# My Ontologist: Evaluating BFO-Based AI for Definition Support*

Carter-Beau Benson†[1,2,3], Alec Sculley[3,4], Austin Liebers[3,4] and John Beverley[2,3]

[1] *CUBRC Inc, Buffalo, NY, USA*
[2] *State University of New York at Buffalo, Buffalo, NY, USA*
[3] *National Center for Ontological Research, Buffalo, NY, USA*
[4] *Summit Knowledge Solutions, Washington, DC, USA*

**Abstract**

The adoption of the Basic Formal Ontology (BFO) and the Common Core Ontologies (CCO) by the U.S. Department of Defense and Intelligence Community has heightened the demand for BFO-conformant ontology development at a pace that far exceeds the availability of trained ontologists. In parallel, large language models (LLMs) such as GPT-4 and GPT-5 have demonstrated striking generative capabilities, raising the prospect that they might accelerate ontology engineering. This paper evaluates that prospect through *My Ontologist*, a custom GPT-based assistant designed to generate BFO-aligned ontology content. Across a series of experiments—from baseline prompting through three iterative system designs, and including the effects of GPT-4o and the shift to opaque IRIs in CCO—we assess the extent to which LLMs can reliably produce Aristotelian definitions, respect BFO/CCO hierarchies, and generate valid Turtle/OWL code. Results show that while models can output syntactically correct triples and occasionally identify proper parent classes, they frequently misclassify entities, invent relations, and deviate from ontological commitments. Explicit constraints, modular ontology resources, and rule-based prompts improved performance in *My Ontologist 3.0*, yet instability remained, and GPT-4o regressions underscored the fragility of relying on closed models. We conclude that LLMs can function as useful assistants in BFO-based ontology work, automating repetitive tasks and drafting candidate content, but they cannot replace trained ontologists. Their outputs must be constrained by formal rules, validated against authoritative ontologies, and reviewed under human oversight. Future work should focus on open-source models fine-tuned for ontology tasks, embedded within validation-driven pipelines, to combine the flexibility of LLMs with the rigor demanded by BFO.

## 1. Introduction

Ontologies and knowledge graphs are increasingly recognized as foundational tools for organizing, integrating, and reasoning over complex data. Their value is especially pronounced in national security and defense, where semantic precision and interoperability are essential for mission-critical decision-making. The Basic Formal Ontology (BFO) and its mid-level extension, the Common Core Ontologies (CCO), have been formally adopted as the baseline standards for ontology development by the Department of Defense (DoD) and the Intelligence Community (IC) [1]. This adoption underscores the importance of upper-level ontologies in ensuring consistent and interoperable representations across agencies and operations.

The widespread adoption of BFO has created a premium on the expertise of trained ontologists, even as their numbers remain limited. Ontology development, moreover, remains a labor-intensive and time-consuming endeavor. The slow pace of manual ontology construction often fails to keep pace with rapidly evolving mission requirements, producing a gap between operational needs and available ontology content. This shortage of trained personnel and the inherent slowness of traditional development cycles compound one another, making it difficult to deliver high-quality, BFO-conformant ontologies at the speed required in defense contexts.

† Corresponding Author, carter-beau.benson@CUBRC.org

In parallel, large language models (LLMs) have emerged as a transformative technology in artificial intelligence. Since the release of GPT-3.5 in 2022, followed by GPT-4, GPT-4o, and most recently GPT-5, these systems have demonstrated unprecedented capabilities in generating coherent text, computer code, and structured information with minimal prompting [2]. Their rapid adoption across research and industry highlights the potential for LLMs to accelerate knowledge-intensive workflows, raising the question of whether they might also help alleviate the bottleneck in ontology engineering.

However, the unconstrained application of LLMs to ontology construction risks introducing significant errors. Naïve automation strategies, such as converting database schemas or spreadsheets directly into ontology classes, often yield poorly structured vocabularies that lack logical rigor or semantic interoperability. LLMs, without principled constraints, risk reproducing these problems at scale: generating ambiguous hierarchies, fabricated relations, and logically inconsistent structures. In defense and intelligence applications, such weaknesses are especially costly, as they undermine the very interoperability that ontologies are intended to secure. Although some recent work has explored LLMs for ontology alignment [3,4], knowledge graph construction [5–7], and conversational ontology engineering [8], few efforts have directly addressed the challenge of generating ontologies that conform to an established upper-level framework such as BFO.

This paper investigates whether LLMs can be effectively constrained to produce ontology content that is both rapid to generate and conformant with BFO. To this end, we developed *My Ontologist*, an experimental system that combines modular BFO and CCO resources with a rule-based definition schema and prompts GPT models to output Turtle/OWL class definitions. We evaluate multiple GPT variants, from 4 through 5, on tasks of definition generation and class hierarchy construction. While constrained prompting improves syntactic correctness and occasionally yields BFO-conformant outputs, persistent challenges remain, including model drift, property invention, and structural brittleness. Our results suggest that while LLMs cannot replace the expertise of trained ontologists, they may serve as valuable assistive tools when embedded within a supervised, validation-enhanced workflow.

## 2. Related Work

Work on applying large language models (LLMs) to ontology engineering can be grouped into ontology alignment, knowledge graph construction, interactive ontology development, ontology learning, and upper-level integration with modular pipelines. Each strand addresses part of the ontology engineering challenge, but none directly resolves the issue of generating ontology content that conforms to upper ontologies such as BFO or CCO.

Research on ontology alignment has explored using deep learning and LLMs to improve cross-ontology matching. He et al. introduced BERTMap, which uses contextual embeddings fine-tuned on ontology pairs, supplemented with structural features, to predict mappings between classes across different vocabularies [3]. Hertling and Paulheim proposed OLaLa, a prompting-based approach that applied LLMs to the Ontology Alignment Evaluation Initiative (OAEI) tasks and compared zero- and few-shot strategies [4]. While both studies report gains over traditional lexical and structural methods, their scope is limited to reconciling existing ontologies rather than generating new content, and they do not enforce constraints from upper ontologies.

A second line of work applies LLMs to knowledge graph construction and population. Caufield et al. developed SPIRES, which uses structured prompting and recursive interrogation of LLMs to extract candidate triples from text for schema-guided KG population [5]. Meyer et al. explored ChatGPT for knowledge graph engineering, documenting its ability to suggest entities and relationships but also its tendency to generate fabricated or inconsistent triples [6]. Carta et al. proposed an iterative zero-shot prompting pipeline, in which extracted triples are refined through multiple LLM passes without task-specific training data [7]. These approaches illustrate the potential of LLMs to accelerate KG construction, but they typically emphasize coverage and efficiency, leaving semantic validity and upper-level alignment to subsequent curation.

Other research frames ontology engineering as an interactive or conversational process. Zhang et al. introduced OntoChat, a system that allows users to iteratively refine ontology content through natural language dialogue with an LLM [8]. The framework supports requirements elicitation and competency questions, lowering the barrier for non-experts to participate in ontology construction. While effective for usability, OntoChat prioritizes interaction quality and flexibility rather than ensuring that the generated ontology content conforms to the commitments of an upper-level ontology.

LLMs have also been tested for ontology learning and class generation. Babaei Giglou et al. evaluated nine LLM families on tasks including term typing, taxonomy induction, and non-taxonomic relation extraction, reporting that models performed better at identifying candidate terms and hierarchies than at capturing complex semantic relations [9]. Lo et al. proposed OLLM, a fine-tuned end-to-end model that generates entire taxonomy backbones from text, using a novel graph-similarity evaluation metric to compare learned structures against reference ontologies [10]. While these studies demonstrate that LLMs can partially automate ontology induction, they do not incorporate upper-level frameworks such as BFO or CCO to constrain generated classes and relations.

More recent work attempts to integrate LLM outputs with formal or upper ontologies. Procko, Ochoa, and Elvira developed a GPT-based pipeline for generating Aristotelian definitions in OWL, anchoring each definition to an asserted BFO parent to enforce genus–differentia form [11]. Toro et al. introduced DRAGON-AI, a retrieval-augmented framework that combines GPT-4 with existing biomedical ontologies and literature to propose new classes, definitions, and axioms [12]. Expert reviewers found that DRAGON-AI achieved high precision in selecting parent classes, but that textual definitions often required revision. Joachimiak et al. reported on the Artificial Intelligence Ontology (AIO), which was developed with LLM-assisted term suggestions and draft definitions under curator supervision [13]. In all three cases, human oversight remained central to ensuring quality, and the methods did not provide systematic safeguards against common modeling errors such as role/material entity confusion.

Finally, several projects have experimented with modular or agent-based pipelines. Wawrzik et al. presented KGG4SE, a systems engineering framework that integrates GPT-4 with iterative OWL reasoning checks to maintain logical consistency against BFO during ontology generation [14]. Sadruddin et al. proposed SCHEMA-MINERpro, which combines schema extraction via GPT-4 with an ontology-grounding agent that maps extracted elements to the QUDT ontology through a combination of lexical heuristics, semantic similarity, and fallback LLM prompts [15]. Both approaches illustrate a shift from monolithic prompt-response models to modular architectures in which LLMs are embedded as components within validation-driven workflows.

Across alignment [3,4], knowledge graph construction [5–7], conversational interfaces [8], ontology learning [9,10], upper-ontology integration [11–13], and modular pipelines [14,15], LLMs have been applied to a wide range of ontology-related tasks. These studies show that LLMs can generate candidate classes, definitions, and relations, and that modular validation mechanisms can improve quality. However, most approaches address semantic correctness and upper-ontology compliance only *after* generation, either through human review or reasoning checks. Some work has applied BFO constraints to definition writing [11], and others incorporate reasoning loops to repair outputs [14], but there is little evidence of systems that systematically enforce BFO or CCO commitments *during* the generation of ontology content. In domains such as defense, where BFO has been formally adopted [1], this limitation is particularly significant. The present study addresses this gap by evaluating whether a tailored GPT-based agent can be guided to produce BFO-conformant ontology content at the point of generation. The goal is not to determine which GPT variant performs best in isolation, but rather to test whether current LLMs, when carefully constrained, can fulfill this role at all—thereby shifting the function of trained ontologists from primary modelers to supervisors and validators.

# 3. Methods and Results

The development of *My Ontologist* proceeded iteratively, with each stage consisting of a methods phase followed by a results phase. We began with a baseline experiment in which GPT-4 was supplied with minimal ontological input to test its ability to generate BFO-conformant definitions. Building on the shortcomings identified there, we implemented three successive versions of *My Ontologist*. Version 1 incorporated a curated set of reference materials but demonstrated significant limitations in producing usable definitions. Version 2 expanded the knowledge base to include full versions of BFO and CCO along with several domain ontologies, resulting in new challenges. Version 3 introduced a codified set of thirty-six rules, modular ontology resources, and representations in both OWL and first-order logic. Each iteration is described in paired methods and results subsections, documenting both the technical adjustments and the outcomes they produced.

# 4. Baseline

## 4.1 Baseline Methods

To generate triples in alignment with a standard, we first provided the necessary annotation property IRIs from a preexisting ontology. For example, we supplied the following properties:

- **Label** – http://www.w3.org/2000/01/rdf-schema#label
- **Subclass** – http://www.w3.org/2000/01/rdf-schema#subClassOf
- **Definition** – http://www.w3.org/2004/02/skos/core#definition
- **Contributor** – http://purl.org/dc/elements/1.1/contributor
- **Example** – http://www.w3.org/2004/02/skos/core#example

These IRIs were placed into a GPT-4 prompt that asked the model to generate a BFO-conformant definition for a given term. The model was required to output the result in Turtle syntax. This setup provided a minimal baseline for evaluating whether GPT-4 could independently produce definitions and axioms consistent with BFO given only annotation IRIs and a target term.

## 4.2 Baseline Results: Simple Case

When asked to define *Human*, GPT-4 produced the following Turtle triples:

```
@prefix rdfs: <http://www.w3.org/2000/01/rdf-schema#> .
@prefix skos: <http://www.w3.org/2004/02/skos/core#> .
@prefix dc: <http://purl.org/dc/elements/1.1/> .
@prefix : <http://example.org/ontology#> .
@prefix bfo: <http://purl.obolibrary.org/obo/BFO_> .
@prefix owl: <http://www.w3.org/2002/07/owl#> .
```

```
:Human rdfs:label "Human" ;
    rdfs:subClassOf bfo:0000040 ; # material entity
    skos:definition "A material entity that is a member of the species Homo sapiens,
      characterized by complex organ systems and capable of cognitive functions and social
      behaviors." ;
    dc:contributor "Jane Doe" ;
    skos:example "John Doe participating in a complex social interaction." .
```

GPT-4 correctly identified *Human* as a subclass of BFO_0000040 (Material Entity), as evidenced by similar modeling in CCO and the Open Biomedical and Biology Ontology Foundry, and provided an Aristotelian-style definition with plausible differentiating features. However, it leveraged concepts unanalyzed in the current iteration of BFO, like "capable" and provided probabilistic differentia that some humans do not satisfy. At a surface level, this output suggested that the model could map a term to the proper superclass and generate a syntactically correct RDF definition.

## 4.3 Baseline Results: Complex Case

When asked to define *Student*, GPT-4 produced the following triples:

```
@prefix rdfs: <http://www.w3.org/2000/01/rdf-schema#> .
@prefix skos: <http://www.w3.org/2004/02/skos/core#> .
@prefix dc: <http://purl.org/dc/elements/1.1/> .
@prefix : <http://example.org/ontology#> .
@prefix bfo: <http://purl.obolibrary.org/obo/> .

:Student rdfs:label "Student" ;
    rdfs:subClassOf bfo:BFO_0000040 ; # Material Entity
    skos:definition "A material entity that is a human, formally engaged in learning, typically
    enrolled in an educational institution such as a school or university." ;
    dc:contributor "John Doe" ;
    skos:example "Jane Doe studying at a university library." .
```

The first and most serious problem is that GPT-4 ontologically misclassified *Student* as a subclass of *Material Entity*. Within the framework of BFO, this is a category mistake with significant consequences. Material entities are defined in BFO as independent continuants characterized by their intrinsic physical constitution. Instances of material entities are what they are by virtue of their material makeup; if they cease to have that constitution, they cease to exist as members of that class. Humans, for example, are modeled as material entities because they are organisms with a specific biological structure, and they remain humans as long as that structure is preserved. Roles, by contrast, are specifically dependent continuants that inhere in material entities. They do not change the underlying physical makeup of the bearer; instead, they reflect contingent, contextual, or institutional states of affairs. A *Student* is not a different kind of organism than a *Non-Student*; it is a human who bears the role of student under specific circumstances, such as being enrolled in an educational program. By treating *Student* as a subclass of *Material Entity*, GPT-4 erased this distinction between intrinsic kinds and contextual roles. This error undermines the logical integrity of any ontology built on such definitions, since it leads to axioms that equate socially conferred statuses with biological kinds. From the standpoint of ontology engineering, this kind of misclassification has cascading effects: it makes reasoning less reliable, confuses queries, and ultimately threatens the interoperability of ontologies across domains that depend on precise role modeling, such as defense, medicine, or education.

The second problem concerns the structure of the definition itself. BFO and its associated ontology community adhere to an Aristotelian schema for definition writing, in which the form is expected to be "An X is a Y that Zs." This pattern serves several critical purposes. It identifies the immediate genus of the class (*Y*), which ensures proper placement in the taxonomic hierarchy, and it specifies the differentia (*Z*), which distinguishes the class from its siblings and clarifies its scope. By following this structure, ontologists can maintain consistency across definitions, reduce cognitive load when reading or evaluating them, and support automated reasoning tools that depend on clear genus–differentia boundaries. GPT-4, however, produced a definition for *Student* that strayed from this schema. Instead of the expected linear form, it generated a nested structure closer to "X is a Y that is a Z that Es." This construction obscures the immediate parent class, since the genus is buried under layers of further qualification, and it muddies the role of the differentia, which becomes spread across multiple clauses. The effect is to diminish clarity and interpretability for both human readers and automated systems. For ontology engineers, the danger is that such definitions resist straightforward translation into axioms: if the genus cannot be easily extracted, then placement in the hierarchy becomes unstable, and if the differentia are not clearly demarcated, then sibling classes cannot be reliably distinguished. Thus, even when GPT-4's output appeared to mimic the Aristotelian style, its departure from the strict form compromised both the usability and the formal rigor of the resulting definition.

A third and equally significant issue lies in the ambiguity of the differentia provided by GPT-4. In BFO, the differentia is the critical element that distinguishes a class from its siblings under a shared genus. It must be precise, unambiguous, and logically sufficient to mark off the scope of the class. GPT-4's proposed differentia for *Student*—phrases such as "formally engaged in learning" and "typically enrolled in an educational institution"—fails to meet these requirements. The expression "formally engaged in learning" introduces unnecessary vagueness. If taken literally, it could imply that the status of being a student is suspended whenever a person is asleep, distracted, or momentarily disengaged, since during those times they are not "formally engaged." Moreover, the term "formal" itself is underspecified. Does it require enrollment in an accredited institution, participation in a recognized curriculum, or simply adherence to a structured process of instruction? Different ontological interpretations are possible, none of which are clarified by GPT-4's definition. The addition of "typically enrolled in an educational institution" exacerbates the problem. Rather than providing a crisp differentiating feature, this phrase functions more as sociological commentary, gesturing toward a statistical generalization rather than an ontological criterion. Such wording is inadequate for formal ontology, where classes must be defined in terms of necessary and sufficient conditions, not in terms of typical or contingent associations. The result is a definition that blurs rather than sharpens the boundary of the class, contradicting BFO's emphasis on clarity and precision. Ambiguity at the level of the differentia introduces risks not only of misclassification but also of interoperability failures, since other ontologies or systems may interpret these terms differently, further fragmenting the semantic landscape.

A final issue, separate from the misclassification itself, is GPT-4's failure to reference the BFO class *Role* as the proper corrective category. While the earlier problem was that the model actively placed *Student* under *Material Entity*, this problem lies in what the model left out. The definition it generated implicitly acknowledged that a student is a human in a particular circumstance, but it never tied that circumstance back to the ontological mechanism BFO provides for handling such cases—namely, roles as specifically dependent continuants that inhere in bearers. This omission matters because referencing *Role* would have signaled an understanding that studenthood is contingent and context-dependent, not intrinsic to the organism itself. By failing to invoke this category, GPT-4's output was not only incorrect in placement but also incomplete in form: it missed the chance to repair its own mistake by situating the definition within BFO's existing framework for contingent statuses. The result is a definition that cannot be reconciled with the upper ontology, because it both misclassifies the entity and neglects to employ the very construct designed to prevent such confusion

## 4.4 Baseline Discussion

The definition of *Student* generated by GPT-4 was not false in the ordinary sense. From a natural language perspective, it is intelligible to describe a student as "a human, formally engaged in learning, typically enrolled in an educational institution." Many readers would accept this as a reasonable characterization, and in an everyday context it would pass as adequate. The problem, however, is not with its truth-value in ordinary discourse but with its failure to meet the standards required by BFO. Ontology construction in the BFO tradition does not aim merely at naturalistic description; it aims at precise, logically coherent representations that can support reasoning and integration across domains. Judged by these standards, GPT-4's output was inadequate.

The missteps in the *Student* definition—misclassification under *Material Entity*, deviation from the Aristotelian schema, ambiguous differentia, and the failure to reference roles—demonstrate that GPT-4 can produce definitions that look acceptable at first glance yet fail to align with BFO's commitments. This distinction between surface plausibility and ontological correctness is critical. Without sensitivity to the formal categories and definition rules that govern upper-level ontologies, LLM-generated content risks introducing subtle errors that propagate throughout an ontology. Such errors may not be immediately obvious, since they do not render the definition incoherent as English prose, but they undermine the logical structure that BFO enforces

The exercise therefore highlights a key insight: large language models are capable of generating definitions that conform to certain stylistic expectations, but they do not reliably enforce the deeper ontological distinctions that make frameworks like BFO effective. The gap between what appears reasonable in natural language and what is valid within BFO is precisely what motivates the use of formal rules and structured oversight in ontology engineering. The baseline results showed that GPT-4 is not unusable—it produced syntactically valid RDF, employed annotation properties correctly, and in simple cases mapped terms to plausible parents—but that its apparent successes mask deeper problems. From a BFO perspective, these problems are not incidental; they are structural, and they call for systematic constraint and validation beyond what the model can provide unaided.

## 4.5 Baseline Conclusion

The baseline experiment provided a useful starting point for understanding both the potential and the limitations of GPT-4 in ontology development. On the positive side, the model demonstrated an ability to generate well-formed Turtle syntax, to apply annotation properties appropriately, and, in simple cases, to identify reasonable superclass relations. These results suggest that LLMs can automate parts of the technical labor traditionally carried out by ontology engineers, such as producing machine-readable triples and populating OWL files. Yet the *Student* example revealed the far more consequential weaknesses. The model could produce definitions that were coherent in ordinary language but that broke with BFO's ontological commitments, introducing errors that would undermine any ontology built from them. This gap underscores a fundamental point: success at the level of language generation does not equate to success at the level of ontological modeling.

The lesson of the baseline is therefore twofold. First, GPT-4 can be a useful assistant for ontology construction if it is given clear structures to follow and if its output is subject to careful human review. Second, unconstrained prompting is insufficient, because it allows surface plausibility to mask violations of formal modeling standards. If LLMs are to be meaningfully incorporated into ontology engineering workflows, their use must be guided by explicit rules and anchored in established ontologies such as BFO and CCO. The baseline results, while mixed, established a clear motivation for moving beyond ad hoc prompting toward a more structured system—what became the first iteration of *My Ontologist*.

# 5. My Ontologist 1.0

## 5.1 Methods

The first version of *My Ontologist* was created with the explicit goal of serving as an expert assistant in generating Basic Formal Ontology (BFO)-conformant ontologies. The design assumption was that if the system were provided with the authoritative texts on BFO and related technical resources, it would be able to use these as guides when producing definitions, subclass hierarchies, and RDF/OWL output. ChatGPT allows for the creation of a custom GPT, where up to twenty PDF files could be uploaded to form the model's private knowledge base. Because GPTs could not directly read .ttl or .owl files, these ontology resources had to be converted to .pdf format in order to be accessible.

The sources chosen for *My Ontologist 1.0* reflected both conceptual and technical needs. The *Basic Formal Ontology 2.0: Specification and User's Guide* [16] was included to anchor the system in the philosophical commitments of BFO, particularly the distinctions between continuant and occurrent, the use of Aristotelian definition form, and the principles of ontological realism, perspectivalism, fallibilism, and granularity. The *BFO 2020 Turtle code*, converted to PDF, was added to give the model direct access to class labels and hierarchy in a structured form [17], thereby ensuring that definitions could be aligned with the official vocabulary and parent classes of BFO. Finally, *Learning SPARQL* [18] was provided as a technical reference, intended to give the model a foundation in query language syntax and the mechanics of generating triples. Together, these resources were intended to allow the GPT to follow the principles stated in the specification, extend from the canonical BFO modules, and generate OWL files that conformed both structurally and semantically to BFO. Prompts were then supplied to the system, with positive or negative feedback given depending on the adequacy of the responses.

## 5.2 Results

The results of this initial setup did not align with expectations. Rather than completing the modeling tasks it was given, *My Ontologist 1.0* often defaulted to explaining how such tasks should be carried out. When asked to model a term like *Student*, the system responded with lengthy discussions of perspectivalism, ontological realism, or granularity in modeling in BFO. On repeated prompting, it shifted to reiterating the difference between *continuant* and *occurrent*, the two highest-level categories in BFO. Only after extended effort and patient redirection could a user coax 1.0 into producing an actual definition, and even then the output was typically no better than what had been generated in the baseline experiment with GPT-4. Definitions broke from established convention, failed to follow the Aristotelian pattern, and did not ask clarifying or disambiguating questions when ambiguity was present.

## 5.3 Discussion and Conclusion

These outcomes highlight a fundamental limitation of the approach used in *My Ontologist 1.0*. Supplying the model with PDFs containing BFO specifications, ontological guides, and technical manuals gave it access to large quantities of descriptive material, but this content was not directly translatable into the practice of generating class definitions. Instead of applying the principles of BFO to the modeling task at hand, the system repeatedly regurgitated explanatory passages from its knowledge base. The long discourses about perspectivalism or continuant/occurrent distinctions were not irrelevant to ontology, but they displaced the actual task of producing BFO-conformant definitions. Our interpretation is that the GPT placed disproportionate weight on instructional and expository materials, particularly the *BFO 2.0 Specification* and the BFO Reference documents, and treated the prompt more as an invitation to elaborate on those resources than as a command to execute.

*My Ontologist 1.0* demonstrated that simply providing an LLM with authoritative ontological texts is insufficient to constrain it to productive modeling behavior. The system showed awareness of BFO principles, but it applied that awareness discursively rather than procedurally. It could talk about ontology at length, but it could not consistently perform ontology-building tasks in alignment with BFO. These results underscored the need to move beyond loading general references into the model's backend and toward approaches that imposed more explicit rules and structures on its behavior. This realization directly motivated the development of *My Ontologist 2.0*.

# 6. My Ontologist 2.0

## 6.1 Methods

*My Ontologist 2.0* was designed to overcome the limitations observed in the first version by providing the model with a richer and more domain-diverse knowledge base. To accomplish this, the backend was expanded to the maximum twenty PDF files permitted. In addition to the BFO specification and reference documents, the entire suite of Common Core Ontologies (CCO) was included, converted to PDF form to make them accessible to the GPT [19]. The *CCO NIST User's Guide* was also supplied, offering concrete examples of how CCO modules are intended to be used in modeling tasks [20]. To test whether domain-specific grounding could improve definitional precision, several publicly available ontologies produced by CUBRC were added, covering the cyber domain and cognitive process domain [19]. Finally, a set of BFO-related academic papers were uploaded to provide philosophical and methodological background on ontological realism and upper ontology usage. The expectation was that this expanded and diversified knowledge base would allow the GPT to move beyond generic explanations and to produce class definitions that were properly aligned with BFO and CCO while remaining sensitive to domain-specific requirements.

## 6.2 Results

The outcomes of this expansion, however, were not encouraging. The inclusion of the entire CCO suite—over a thousand pages in length—proved to be a liability. Instead of facilitating targeted modeling, the size and complexity of the document often led the GPT to select inappropriate parents for classes, creating definitions that deviated from CCO's established taxonomic structure. Many of the same problems observed in version 1 persisted: definitions remained verbose, Aristotelian structure was inconsistently applied, and no disambiguation questions were asked. Worse still, the system frequently crashed before completing a definition, unable to process or retain coherence across the length of the massive PDFs. The result was a model that was slower, less reliable, and in some ways even less effective than the baseline or *My Ontologist 1.0*.

## 6.3 Discussion and Conclusion

The poor performance of *My Ontologist 2.0* illustrates that simply expanding the range of reference materials does not translate into better definitional modeling. On the contrary, the attempt to load the entire CCO suite and multiple domain ontologies overwhelmed the system. Rather than producing carefully constrained definitions, the GPT was swamped by the volume of material and defaulted either to generating generic explanatory text or to collapsing mid-response. This suggests that effective use of LLMs for ontology engineering cannot rely on brute-force access to large knowledge bases, even when those knowledge bases are themselves authoritative. Instead, models must be guided by targeted constraints and lightweight resources that prioritize clarity and structure over exhaustiveness.

*My Ontologist 2.0* confirmed that expanding the scope of backend knowledge is not sufficient to improve ontology generation. The model's outputs showed no consistent improvement over *My Ontologist 1.0* and, in many cases, degraded in quality due to the unwieldy size of the resources provided. These results underscored the need for a different approach: one that would abandon the strategy of overloading the GPT with ontological texts and instead impose explicit, codified rules of definition writing and ontology construction. This realization provided the foundation for *My Ontologist 3.0*.

# 7. My Ontologist 3.0

## 7.1 Methods

The lessons learned through the design failures of 1.0 and 2.0 were instrumental in creating a more stable and informative *My Ontologist 3.0*. Rather than providing academic papers or massive PDFs, 3.0 was given thirty-six rules that offered explicit guidance for the creation of ontology classes. These rules emphasized clarity, objectivity, coherence, extensibility, user engagement, and the precise articulation of concepts. Among the directives were requirements to prefer defined classes over primitive classes, to adhere strictly to the Aristotelian definition schema, to refrain from creating new properties, and to adopt consistent naming conventions—such as appending "role" to subclasses of *Role* and "act" to processes, for example "*x* role" and "act of *y*". A full list of these rules is available upon request.

In terms of resources, *My Ontologist 3.0* was supplied with *BFO 2020* and the entirety of the *Common Core Ontologies (CCO)* suite, but this time in modular rather than monolithic form. Individual modules included the Quality Ontology, Agent Ontology, Extended Relations Ontology, Artifact Ontology, Time Ontology, Geospatial Ontology, Currency Ontology, Event Ontology, Units of Measure Ontology, Facility Ontology, and Information Entity Ontology [19]. Finally, to overcome the limitations of OWL's description logic expressivity, *My Ontologist 3.0* was given access to first-order logic (FOL) formulations of BFO in Common Logic Interchange Format (CLIF) [17]. This inclusion was intended to allow the GPT to better capture the logical structures of BFO, recognizing that OWL enforces decidability at the cost of expressive richness.

## 7.2 Results

Compared to its predecessors, *My Ontologist 3.0* showed significant improvement. When prompted with an ambiguous term, it now asked clarifying questions to disambiguate before generating a definition. For example, it modeled *Student* not as a material entity but as a role borne by a human, labeling it “Student Role” and providing conditions under which it is realized. Its generated definition was as follows:

```
:StudentRole rdf:type owl:Class ;
        rdfs:subClassOf bfo:0000023 ; # Role
        rdfs:label "Student Role" ;
        skos:definition "A role that is realized by an individual who is enrolled in and actively participating as a learner in educational processes."@en ;
        rdfs:comment "This role applies to individuals who are registered students in educational courses or programs."@en .
```

Here, *My Ontologist 3.0* correctly identified the parent class as *Role* and named the entity accordingly, showing progress over both the baseline and version 1.0. However, it used a BFO relation “realized” in an unconventional way. Roles are realized by process. Here, it appears that the role is realized by an individual, although it could be interpreted that the realization is the educational process. Regardless, the proper property is “inheres in”. As was the case before, *My Ontologist* provided problematic differential. That said, the recognition that an analysis of *Student* required a role was an improvement.

Further evidence of progress can be seen in an extended exchange over the term *Vehicle Bed Mount*. The system’s first attempt described the entity as “affixed” to a vehicle bed introducing an object property not found in CCO. On its second attempt, it introduced a nonexistent class, “artifact component.” On the third and fourth tries, it again used object properties not sanctioned by CCO. Only on the fifth attempt did it succeed:

> “Vehicle Bed Mount is an equipment mount that is part of the vehicle bed, designed to secure and stabilize equipment or cargo during transport.”

This final response extended a valid CCO class (*equipment mount*) and used a legitimate BFO object property (*part of*). Notably, the definition converged closely on an official CUBRC definition—“Vehicle Bed Mount is an equipment mount that is designed to attach to the bed of a vehicle”—despite *My Ontologist 3.0* not having access to that reference.

However, failures remained. In a lengthy series of prompts around the term *Act of Booking*, the system consistently struggled to identify a proper parent class. It produced candidates such as “act of commitment communication,” “act of detaining,” “act of processing,” and “act of custody,” none of which exist in CCO. In one exchange, after repeated corrections, the system finally suggested “Legal System Act” as a parent—a valid class—but then failed to identify a proper object property. When prompted to add one, it invented *involves* and incorrectly assigned it the IRI for *has participant*. Upon correction, the model crashed and had to be reloaded, after which it finally returned:

> “Act of Booking is a Legal System Act that has output in the form of recorded personal and criminal details of a detained individual, processed and documented at a police station by law enforcement officials.”

While this final attempt did use an appropriate CCO class and property, the number of failed intermediate steps, the repeated invention of classes and properties, and the crash itself demonstrated the instability of the system even in its improved form.

## 7.3 Discussion and Conclusion

The performance of *My Ontologist 3.0* underscores the promise and the fragility of LLM-assisted ontology engineering. On one hand, this version showed that explicit procedural constraints and modularized resources could significantly improve model behavior. By supplying a clear set of rules and a streamlined knowledge base, the GPT was better able to align its outputs with BFO's modeling commitments and to avoid the digressive, expository behavior that plagued earlier iterations. The ability to ask disambiguation questions, in particular, was a notable advance, since it signaled that the model could engage in the kind of clarification dialogue that human ontology engineers rely on to reduce ambiguity.

On the other hand, the persistent tendency to fabricate properties, misplace classes, and violate definitional conventions demonstrates that even when heavily constrained, GPT remains unreliable as a stand-alone ontology engineer. The sharp decline in performance following the release of GPT-4o further highlights the risks of building critical workflows on closed, evolving models whose behavior can change without warning. What appeared to be progress toward reliable BFO-conformant modeling was revealed to be contingent and unstable, dependent as much on model versioning as on the rules and resources supplied.

*My Ontologist 3.0* demonstrated the value of explicit constraints and modular resources but also exposed the fragility of LLM-based ontology engineering. The system showed that GPT could be nudged toward BFO-conformant definitions when given carefully crafted rules and curated resources, yet its improvements were partial and precarious. The fabrication of properties and the instability introduced by GPT-4o reveal that LLMs cannot yet be relied upon for consistent, autonomous ontology construction. Instead, they may be best deployed as assistants whose outputs must be validated against formal rules and reviewed by trained ontologists. These lessons motivated further experimentation with integrating external validation layers and pointed toward future directions for combining LLMs with more stable, open-source models.

## 8 GPT-4o and Opaque IRIs

When GPT-4o was first released, *My Ontologist* performed poorly. It ignored backend rules, fabricated properties, and treated its own inventions as authoritative ontology content. These regressions were concerning, but after a period of retuning, performance stabilized and the worst behaviors diminished. In fact, for many tasks GPT-4o became as usable as GPT-4, demonstrating that the model itself was not inherently incapable of contributing to ontology work.

The larger and more enduring challenge came not from the model update but from the shift in the Common Core Ontologies (CCO) to opaque IRIs. Unlike the earlier practice of using semantically informative identifiers, the new IRIs are alphanumeric strings without cues to their meaning. While opaque IRIs are widely considered best practice in ontology engineering—for stability, long-term maintainability, and independence from linguistic drift—they exposed the weaknesses of *My Ontologist*.

With opaque IRIs, *My Ontologist* in GPT- 5 became what might be described as a "lazy but convincing liar." It often produces identifiers that looked very much like authentic CCO IRIs but are, in fact, fabrications. Definitions followed a similar pattern: they mimicked the form and style of CCO definitions but subtly diverged in substance, leaving out essential constraints or adding extraneous commentary. The model did not consult the ontology files until explicitly forced to do so, preferring instead to generate content that appeared plausible. For example, when asked for the definition of "vehicle", my ontologist stated that "vehicle" is not in CCO, found the definition of "conveyance artifact function”, then crafted a definition of “vehicle” using “conveyance artifact function” convincingly in the definition.

Furthermore, several consistent patterns have emerged. First, the model assumed that CCO IRIs were sequentially assigned by kind, when in reality they are alphabetically ordered. This mistake led to systematic misplacements in definitions and hierarchies. Second, the model persistently refused to search for classes by their labels, defaulting instead to guesses based on its own internal patterns. The result was that it generated IRIs and definitions that were close enough to appear correct, but not actually valid within CCO. Only after repeated correction and explicit prompting to consult the ontology files would it acknowledge and resolve these errors.

These behaviors highlight the structural gap between LLMs and ontology best practices. Opaque IRIs, while essential for human-led ontology development, remove the linguistic handles that LLMs rely on for pattern matching. Deprived of these cues, the model defaulted to surface plausibility rather than validation. The result is a system that looks authoritative but cannot be trusted without continuous oversight.

## 9 Conclusion

This paper has documented a series of experiments with *My Ontologist*, a custom GPT-based system designed to assist in the creation of BFO- and CCO-conformant ontologies. Beginning with a simple baseline in which GPT-4 was asked to generate definitions from annotation properties alone, and continuing through three successive versions of *My Ontologist*, we tested how far large language models can be pushed toward reliable ontology construction. Each iteration exposed both strengths and weaknesses. The baseline demonstrated GPT’s ability to produce syntactically correct RDF and plausible definitions but also revealed its tendency to misclassify roles and violate definitional conventions. *My Ontologist 1.0* showed that supplying academic and technical references encouraged explanation rather than modeling. *My Ontologist 2.0*, which attempted to ground the model in the full CCO suite and multiple domain ontologies, collapsed under the weight of its own knowledge base. The breakthrough of *My Ontologist 3.0* came through explicit rules, modular ontology files, and the use of FOL formulations, which improved behavior, led to correct role modeling, and enabled closer alignment with CCO—even if multiple corrections were still often required.

The transition to GPT-4o further demonstrated the fragility of this approach. Although initial regressions threatened to undo earlier progress, retuning eventually restored much of the functionality. Yet the adoption of opaque IRIs in CCO 2.0 introduced a new and more subtle challenge: the system became a “lazy but convincing liar,” generating plausible but fabricated IRIs and definitions unless explicitly forced to consult the ontology files. This revealed that while LLMs can approximate BFO- and CCO-style outputs, they are not naturally oriented toward verification or formal consistency. Their strength lies in generating text that looks right, not in checking whether it is right.

Taken together, these findings suggest a cautious path forward. LLMs can serve as useful assistants for ontology engineering, particularly in automating repetitive tasks such as producing candidate triples, generating draft definitions, or suggesting parent classes. But they cannot yet replace trained ontologists. Their outputs must be constrained by explicit rules, validated against authoritative ontologies, and corrected through human oversight. Moreover, reliance on closed, proprietary models introduces risks of instability, as seen in the regression under GPT-4o, and raises concerns about reproducibility. Future work should explore open-source LLMs fine-tuned on curated

BFO- and CCO-aligned corpora, coupled with external validation layers such as SHACL or SPARQL-based rule-checking, to enforce ontological correctness.

Ontology development in mission-critical domains such as defense and national security is too important to entrust to black-box systems. The promise of *My Ontologist* lies not in demonstrating that GPT can autonomously build ontologies, but in showing how structured constraints, modular resources, and human oversight can channel the generative capabilities of LLMs into tools that support ontology engineers. The experiments documented here highlight both the potential and the limitations of this approach, providing a foundation for future systems that combine the flexibility of LLMs with the rigor of formal ontology.